\documentclass{aa}
\usepackage{graphicx}
\usepackage{txfonts}
\begin{document}
\title{W Hya through the eye of Odin 
\thanks{Based on observations with Odin, a Swedish-led satellite project
funded jointly by the Swedish National Space Board (SNSB), the Canadian 
Space Agency (CSA), the National Technology Agency of Finland (Tekes), and
Centre National d'\'{E}tudes Spatiales (CNES). The Swedish Space Corporation
was the industrial prime contractor and is operating Odin.}
}
\subtitle{Satellite observations of circumstellar submillimetre H$_2$O line emission}
\author{K. Justtanont \inst{1} \and P. Bergman \inst{2} \and B. Larsson \inst{1} 
\and H. Olofsson \inst{1} \and F.L. Sch\"{o}ier\inst{1}
\and U. Frisk \inst{3} \and T. Hasegawa \inst{4} \and \AA~Hjalmarson\inst{2} 
\and S. Kwok \inst{4,5} \and M. Olberg \inst{2} \and Aa. Sandqvist \inst{1}
\and K. Volk \inst{4,6} \and M. Elitzur \inst{7}}

\offprints{K. Justtanont}

\institute{Stockholm Observatory, AlbaNova University Center, SE-106 91 
Stockholm, Sweden
\and Onsala Space Observatory, SE-439 92 Onsala, Sweden
\and Swedish Space Corporation, PO Box 4207, SE-171 04 Solna, Sweden
\and Dept. of Physics and Astronomy, University of Calgary, 
500 University Drive NW, Calgary, Alberta, T2N 1N4, Canada
\and Institute of Astronomy and Astrophysics, Academia Sinica,
P.O.Box 23-141, Taipei 106, Taiwan, R.O.C.
\and Gemini Observatory, Southern Operation Center, La Serena, Chile
\and Department of Physics and Astronomy, University of Kentucky,
Lexington, KY 40506, USA
}

\date{Received 3/3/05; accepted 4/5/05}

\abstract{
We present Odin observations of the AGB star W Hya in the
ground-state transition of ortho-H$_{2}$O, 1$_{10}-1_{01}$, at 557~GHz.
The line is clearly of circumstellar origin. 
Radiative transfer modelling of the water lines observed by
Odin and ISO results in a
mass-loss rate of (2.5$\pm$0.5)\,10$^{-7}$ M$_{\odot}$ yr$^{-1}$, and
a circumstellar H$_{2}$O 
abundance of (2.0$\pm$1.0)\,10$^{-3}$.
The inferred mass-loss rate is consistent with that obtained from
modelling the circumstellar CO radio line emission, and also
with that obtained from the dust emission modelling combined with a 
dynamical model for the outflow.
The very high water abundance, higher than the cosmic oxygen
abundance, can be explained by invoking an injection
of excess water from evaporating icy bodies in the system. The required
extra mass of water is quite small, on the order of $\sim$
0.1\,M$_{\oplus}$.
\keywords{Stars: circumstellar matter -- Stars: evolution --  
Stars: individual : W Hya -- Stars: late-type
-- Stars: mass-loss -- Infrared:stars}
}
\maketitle

\section{Introduction}

One of the main characteristics of an Asymptotic Giant Branch (AGB) star
is its cool, low-velocity outflow. In this outflow, molecules are formed
according to the relative chemical abundance of carbon and oxygen 
(Olofsson \cite{olof03}; Millar \cite{millar} and the references therein).
In addition to H$_2$, CO is 
an abundant species irrespective of the C/O ratio, but in C/O\,$<$\,1 (O-rich) 
objects its abundance is
rivaled by that of H$_2$O, while in C/O\,$>$\,1 (C-rich) objects
HCN and C$_2$H$_2$ also have sizable abundances. Likewise, there are dust
signatures of amorphous carbon and SiC when C/O is larger than unity,
(Treffers \& Cohen \cite{treffers})
while silicates are seen if the reverse is the case (Gilman \cite{gilman}).

Early AGB stars tend to be semiregular variables
with a low mass-loss rate, and these evolve into long-period Mira-type variables
where mass loss is more pronounced. The lifetime of a mass-losing AGB
star is estimated to be about 10$^{5}$ yr, and before evolving off the AGB the
star undergoes an intense epoch of mass loss 
when most of the initial mass is lost, leaving behind a hot C/O core
which then ionizes the surrounding material resulting in a planetary 
nebula (see e.g., Iben \& Renzini \cite{iben}; Habing \cite{habing}).

In this paper we discuss circumstellar H$_2$O observations of the well-studied
O-rich semiregular variable W~Hya, with a period of 382 days (Lebzelter et al.
\cite{leb}).
It has been detected in a number of circumstellar molecular line emissions, 
e.g., CO (Wannier
\& Sahai \cite{wannier}; Loup et al. \cite{loup}; Cernicharo et al.
\cite{cerni};
Kerschbuam \& Olofsson \cite{franz}), SiO maser (Pardo et al. \cite{pardo})
and OH maser lines (Szymczak et al. \cite{szym};
Chapman et al. \cite{chapman}; Etoka et al. \cite{etoka}). It was also detected
in water lines using the Infrared Space Observatory's
(ISO) Short- and Long Wavelength Spectrometers (SWS and LWS)
by Neufeld et al. (\cite{neufeld}) and Barlow et al. (\cite{barlow}),
respectively. Recently, the Submillimeter Wave Astronomy Satellite (SWAS) 
detected the ground-state ortho-H$_{2}$O transition 
at 557~GHz, 1$_{10}-1_{01}$, as reported by Harwit \& Bergin (\cite{harwit}). 

O-rich semiregular variable stars are seen to have multiple 
dust species condensing in their outflows (Cami \cite{cami}). The 10$\mu$m silicate 
feature, dominant in the spectra of Mira variables, may not be the strongest 
emission feature.
This is clearly seen in the spectral energy distribution (SED) of W Hya. 
By fitting the SED with multiple dust species, Justtanont
et al. (\cite{kay04}) derived a dust mass-loss rate of 3\,10$^{-10}$
M$_{\odot}$ yr$^{-1}$ (this is a lower limit as the
13$\mu$m dust feature was not fitted due to the lack of a suitable
identification). This suggests that W Hya is a low-mass-loss-rate AGB star.
However, its dust envelope is known to be very large,
extending out to 40$^{\prime}$ (Hawkins \cite{hawkins}).

\section{Observations}

\begin{figure}
\resizebox{\hsize}{!}{\includegraphics{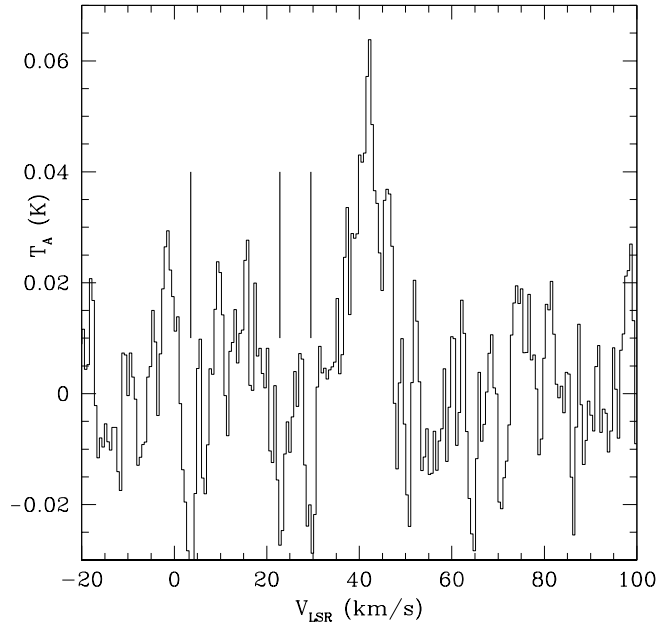}}
\caption{The ground-state transition of ortho-H$_2$O, 1$_{10}-1_{01}$,
as detected by Odin towards W Hya. The three thick vertical lines mark
where the two correlator bands overlap and end.
}
\label{odin_obs}
\end{figure}

The spectrum of the H$_{2}$O (1$_{10}-1_{01}$) line at 557~GHz ($\lambda$ = 
538.29$\mu$m)
was observed by the Odin satellite towards W~Hya using the autocorrelation 
spectrometer (Nordh et al. \cite{nordh}; Frisk et al. \cite{frisk}).
The resulting spectrum is a combination of two separate
observing sessions, in December 2002 and July 2003,
of a total of 185 orbits. As the position-switching mode was used in these 
observations, the on-source integration time
is 53.4 hours. The average system temperature for this
frequency is 3\,200~K.  The calibration was done by switching between 
the internal hot load and the sky. The frequency resolution is 
1 MHz (= 0.54 km s$^{-1}$ at 557 GHz). A sinusoidal baseline was
subtracted and the data rebinned. 
Fig.~\ref{odin_obs} shows the spectrum in terms of
the observed antenna
temperature, $T_{\rm A}$, which has not been corrected for the 
beam efficiency of 90\%. 

The peak antenna 
temperature is 0.05\,$\pm$\,0.02 K. 
The line flux is 
estimated to be 
(3.6$\pm$1.2)\,10$^{-21}$ W\,cm$^{-2}$, after correcting for the
aperture efficiency of 70\% (Frisk et al. \cite{frisk}). 
The absolute uncertainty
in the line flux is large as the data are noisy, and close to the
line there are features which are the artefacts caused by
the ends and the joining of two correlator bands. 
These are seen as absorption features at V$_{\rm LSR}$ of
4, 23 and 30 km s$^{-1}$ (Fig~\ref{odin_obs}).
The estimated uncertainty includes
that due to the noise and the sinusoidal baseline 
subtraction, plus
an absolute flux calibration uncertainty of 10\%. The latter is
obtained by
comparing the line fluxes of Odin and SWAS data for a number of
sources (Hjalmarson 
et al. \cite{ake}). The line intensity is consistent with that obtained by 
SWAS after scaling the latter with the ratio of the Odin and 
SWAS beam sizes, 3.37
(Hjalmarson et al. \cite{ake}).

The centre velocity and the expansion 
velocity  are estimated to be 
42\,$\pm$\,1.5 km\,s$^{-1}$ (LSR scale)
and 7.0\,$\pm$\,1.0 km\,s$^{-1}$, respectively. These values
are consistent with the results from CO 
{\it J}=1--0 data (Kerschbaum \& Olofsson \cite{franz}),
{\it J}=2--1 data (Cernicharo et al. \cite{cerni}), 
{\it J}=3--2 and 4--3 data (Young \cite{young}).

\section{Modelling of the molecular line emission}

In this section, we present our models for both the
H$_{2}$O and CO line emission from the circumstellar envelope of W Hya. 
The models used are based on the assumptions of a constant, isotropic mass-loss rate, 
and a constant expansion velocity, which results in a spherically 
symmetric gas outflow with an $r^{-2}$ density distribution.
The mass-loss rate
refers to the H$_{2}$ loss rate which is the quantity used throughout
this paper. This value needs to be multiplied by
a factor of 1.4 to get the total gas mass-loss rate, due mainly 
to the presence of He.

In order to model the radiative transfer, we need to know the
stellar luminosity. From the spectral energy distribution (SED), the
peak of the spectrum corresponds to an effective temperature
of 2\,500\,K (Justtanont et al. \cite{kay04}), which agrees well 
with that derived by Haniff et al. (\cite{haniff}). 
The Hipparcos distance, taken directly from the Hipparcos
Catalogue (ESA \cite{esa}) parallax, is 115 pc. The
derived luminosity for this distance turns out to be very high --
1.16\,10$^{4}$ L$_{\odot}$ (estimated from a fit to the SED). However, 
the Hipparcos data
were reanalysed by Knapp et al. (\cite{knapp}). The revised 
distance of 78 pc, results in a luminosity of 5\,400 L$_{\odot}$,
more in line with it being an early-AGB star.
The resulting radius is 2.73\,10$^{13}$ cm.
From this point on, we will assume a distance of 78 pc for 
our calculations. 

\subsection{CO lines}

The code used to calculate the CO line emission is
based on the work by Sch\"{o}ier \& Olofsson (\cite{fredrik}).
It employs the Monte-Carlo method to determine the 
excitation in the lowest 40 rotational levels in the
ground and first excited vibrational states. The collisional rates
are taken from the CO-H$_{2}$ rates by Flower (\cite{flower}),
and extrapolated to higher rotational levels and temperatures as described in 
Sch\"oier et al.\ (\cite{schoeier05})\footnote{The molecular 
datafile is available for download at {\tt www.strw.leidenuniv.nl/$\sim$moldata}.}.
The rates for ortho- and para-H$_2$ collisions are weighted assuming an
ortho/para ratio of three.
The outer radius is defined by the
photodissociation of CO, and it is obtained
following the results of Mamon et al. (\cite{mamon}). In order to convert the CO
gas mass-loss rate to the total gas mass-loss rate, we assume a CO
abundance, [CO]/[H$_2$], of 2\,10$^{-4}$.

In the fitting of
observed line fluxes of CO {\it J}=1--0 up to 4--3,
along with the ISO-LWS {\it J}=16--15 and 17--16 lines,
the gas kinetic temperature is calculated with a dust-drag
heating balanced by the adiabatic and radiative
cooling. The radiative cooling is provided by both CO and H$_2$O molecules.
The dust parameters entering in the heating term, due to collisions between
dust and gas particles, are taken to be consistent with those obtained from
the dust modelling. In the case of a low-mass-loss-rate object such as W~Hya,
the CO molecules are mainly excited by the stellar radiation field and the 
temperature structure plays only a minor role (Sch\"{o}ier \& Olofsson 
\cite{fredrik}). A best-fit model is obtained in terms of integrated 
line intensities and in the $\chi^2$ sense,
\begin{equation}
\label{chi2_eq}
\chi^2 = \sum^{N}_{i=1} \frac{(I_{\mathrm{mod},i} -
I_{\mathrm{obs},i})^2}{\sigma^2_{i}},
\end{equation}
where $I$ is the total integrated line intensity, $\sigma_i$ is the uncertainty
in observation $i$, and the summation is done over all independent
observations $N$. The errors in the observed radio line intensities are 
assumed to be 20\%, while the ISO lines have an uncertainty of 30\%.
The result is a stellar mass-loss rate of 1.3\,10$^{-7}$ M$_{\odot}$ yr$^{-1}$.

\begin{figure}
\resizebox{\hsize}{!}{\includegraphics{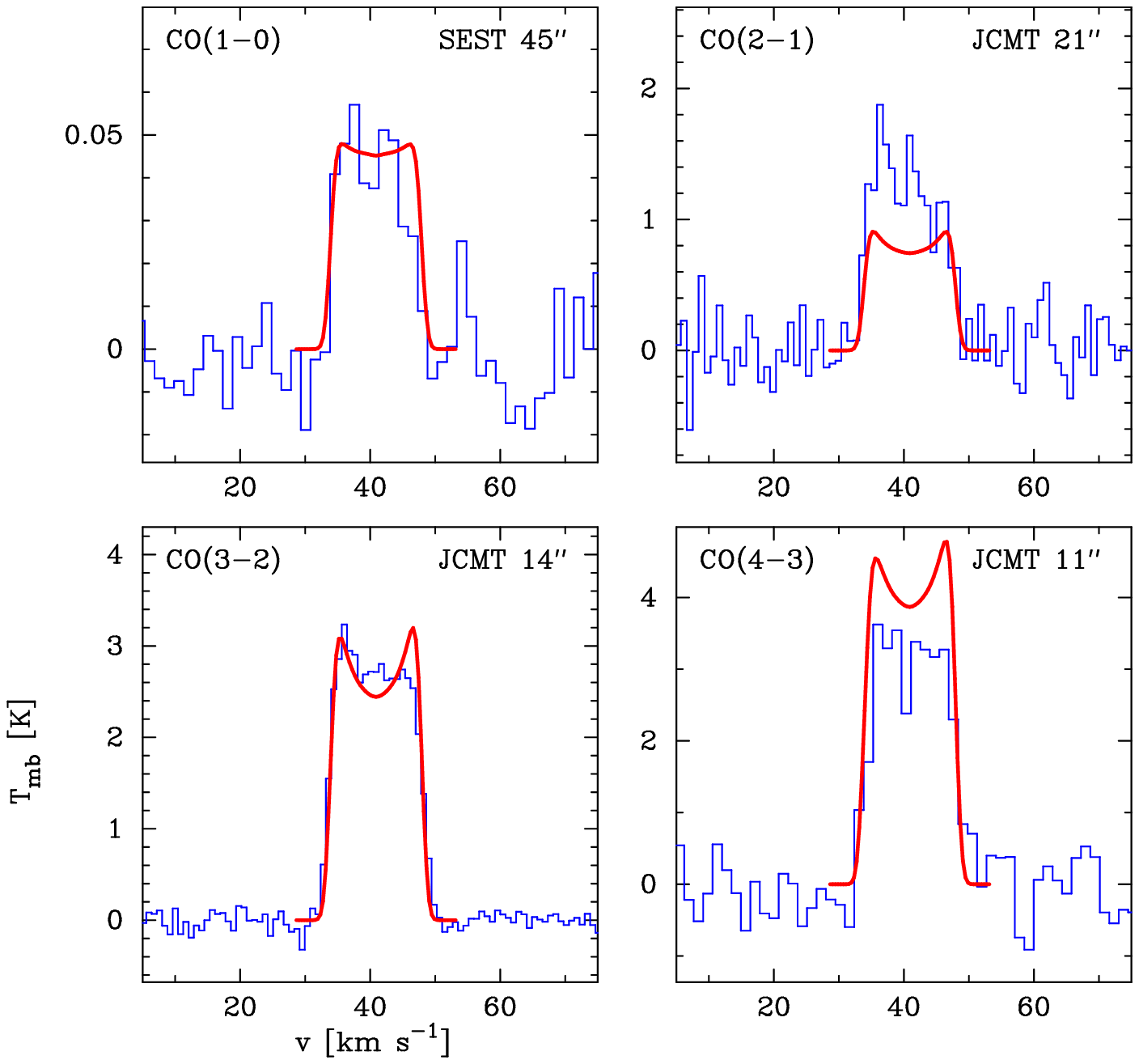}}
\caption{The calculated best-fit model line profiles for CO {\it J}=1--0, 2--1, 3--2,
and 4--3 (continuous line) and the observed profile (histogram).
}
\label{fig_co}
\end{figure}

\begin{figure}
\resizebox{\hsize}{!}{\includegraphics[angle=-90]{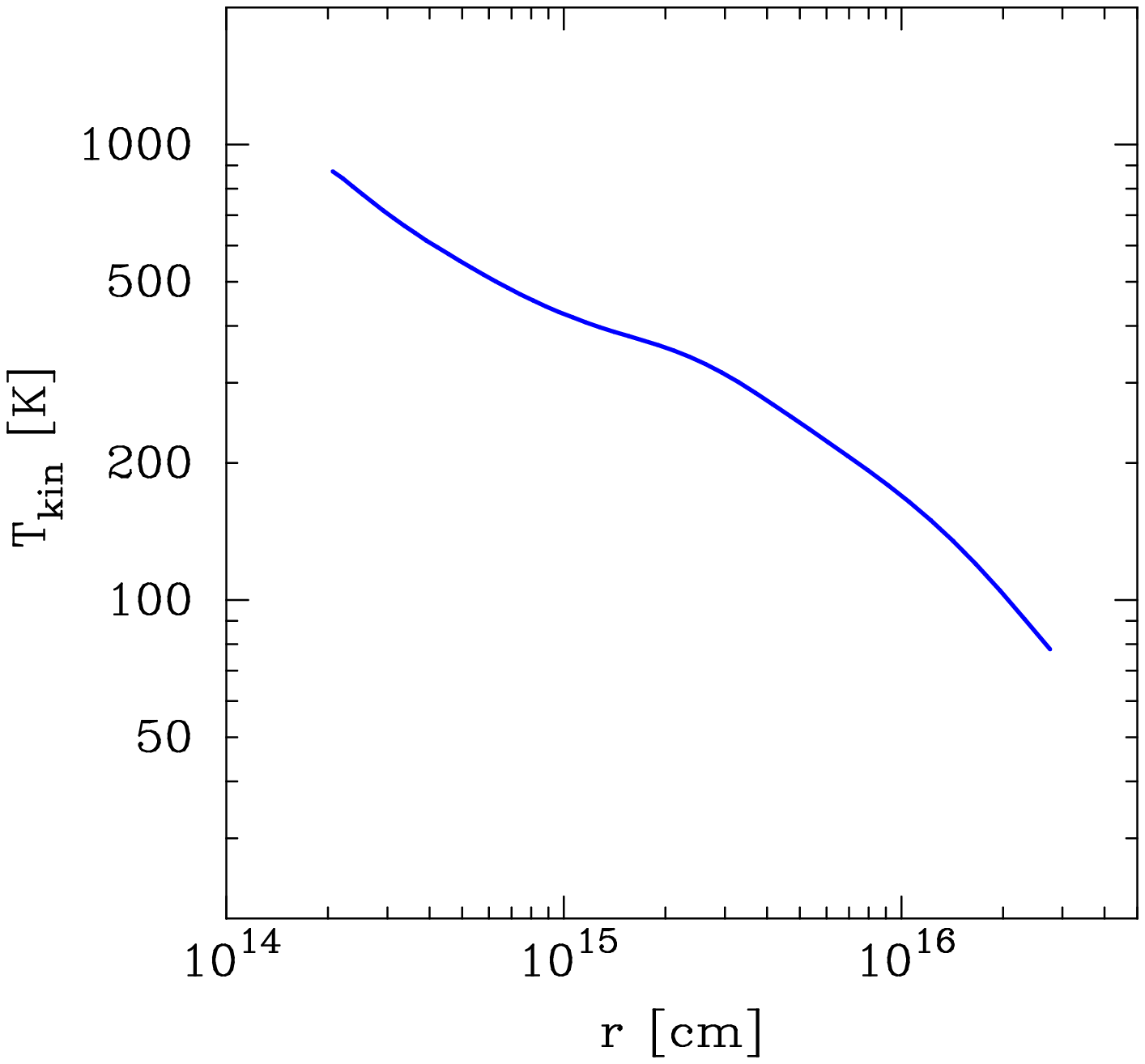}}
\caption{Kinetic temperature profile for the circumstellar envelope of
W Hya derived from our best-fit CO model.}
\label{fig_temp}
\end{figure}

Although our calculated line fluxes are close to the observed,
the line profiles do not match.
We consistently expect to see double-peaked profiles while
the observations tell us otherwise. A way to remedy this is to 
decrease the envelope size and increase the input mass-loss rate.
A mass-loss rate of 1.3\,10$^{-7}$\,M$_{\odot}$\,yr$^{-1}$ results in a 
CO photodissociation radius of 1.4\,10$^{16}$ cm
according to Mamon et al. (\cite{mamon}).
However, our best-fit model, taking into account also the line shapes,
has an outer CO radius of 8\,10$^{15}$ cm with a
mass-loss rate of 2\,10$^{-7}$ M$_{\odot}$ yr$^{-1}$.
The results of the best-fit model are shown in Fig.~\ref{fig_co} and the
line fluxes are listed in Table~\ref{tab_co}. The uncertainty in the
mass-loss rate from fitting the line profile is of the order of 50\% 
within the adopted circumstellar model
(see Sch\"{o}ier \& Olofsson \cite{fredrik}).
The final
model also gives the LSR velocity for W Hya as 40.7 
km\,s$^{-1}$, and an envelope gas expansion velocity of 7.0 km\,s$^{-1}$. 
These values are consistent with those derived from the Odin H$_2$O data.

The final kinetic temperature profile is shown in Fig.~\ref{fig_temp}. 
We note that our derived temperature is between 100--150~K higher than 
that obtained by Zubko \& Elitzur (\cite{zubko}) in the region where the 
H$_2$O molecules are located ($r\lesssim2\,10^{15}$~cm). This is not
inconsistent considering our lower mass-loss rate.

\begin{table}
\caption{Modelled line fluxes for CO compared with the observations
(the unit is K\,km\,s$^{-1}$, except for the ISO lines where the unit
is W\,cm$^{-2}$).
The observed data are taken from the following references :  
a - Kerschbaum \& Olofsson (\cite{franz}), 
b - JCMT archive, c - Cernicharo et al. (\cite{cerni}),
d - Olofsson, private communication,
e - Young (\cite{young}), f - Barlow et al. (\cite{barlow}) }
\begin{tabular}{clccc}
\hline \hline
Transition & Instr. & Observed flux & Model flux & Ref \\
\hline 
1-0   & SEST & 0.82  & 0.66 & a \\
2-1   & JCMT & 18.7  & 11.8 & b \\
2-1   & IRAM & 22.0  & 26.2 & c \\
3-2   & JCMT & 40.2  & 39.8 & d \\
3-2   & CSO  & 29.0  & 22.7 & e \\
4-3   & JCMT & 44.6  & 60.8 & b \\
4-3   & CSO  & 40.0  & 34.8 & e \\
16-15 & LWS  & 3.3\,10$^{-20}$  & 2.4\,10$^{-20}$ & f \\
17-16 & LWS  & 2.3\,10$^{-20}$  & 2.4\,10$^{-20}$ & f \\
\hline
\label{tab_co}
\end{tabular}
\end{table}

\subsection{H$_{2}$O lines}

We model the water lines
observed by Odin, ISO-SWS (Neufeld et al. \cite{neufeld}), and ISO-LWS
(Barlow et al. \cite{barlow}). The code used for this differs from the
code used for CO. The main reason for a different treatment is the extremely
high optical depths in the water lines preventing the convergence
of the Monte-Carlo calculation. The technique employed here is
the accelerated lambda iteration (Rybicki \& Hummer \cite{rybicki}), 
which is used to calculate the excitation in the 45 lowest
rotational levels of ortho- and para-H$_{2}$O.

We use the collisional rates calculated
for H$_{2}$O and He (Green et al. \cite{green}),
with a scaling factor of 1.4 to
correct for the mass difference between H$_{2}$ and He. However,
a more recent calculation for the collision between H$_{2}$O and
ortho- and para-H$_{2}$ by Phillips et al. (\cite{phillips}) suggests
that adopting the collisional rate with He and using the simple scaling
factor can give very different rates. Unfortunately, the
collisional rates with H$_{2}$ were calculated up to 140~K,
while the temperature in the circumstellar envelope of an
AGB star can be up to 1\,000~K. If the H$_{2}$O--He rate is underestimated
by a large factor, it may have a repercussion on the calculated
line fluxes as the water lines are sub-thermally excited
in the envelope of W Hya.

The envelope gas expansion velocity, and the 
kinetic temperature law used are those obtained from the CO line 
modelling.
The outer radius of the water 
molecules is taken from the measurements of the OH
masers by Szymczak et al. (\cite{szym}), who give an outer
OH shell radius of 1.3$^{\prime\prime}$, i.e., 1.5\,10$^{15}$ cm at
the assumed distance. 

The mass-loss rate and the ortho-H$_2$O abundance
are free parameters in the fitting procedure where
the integrated line intensities of the Odin and ISO lines are
used as observational constraints to find the best-fit model
in the $\chi^2$ sense. The observed integrated
line intensities and their uncertainties  are given in 
Table~\ref{tab:models}. We obtained ISO 
fluxes by fitting a Gaussian to archived data with the latest
calibration file (OLP version 10), rather than using the line
fluxes from Barlow et al. (\cite{barlow}) and Neufeld et al.
(\cite{neufeld}). The error in the observed flux reflects only
the estimation of the baseline and does not include possible
systematic error or that of the absolute flux.

\begin{figure}
\resizebox{\hsize}{!}{\includegraphics{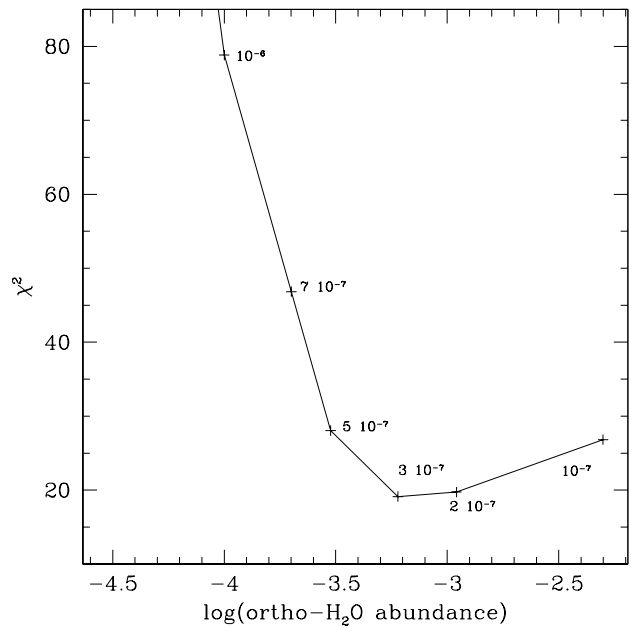}}
\caption{The minimum $\chi^{2}$ for a given combination of
ortho-H$_2$O abundance and stellar mass-loss rate (the number at each
point). The Odin and ISO 
LWS/SWS water line fluxes have been used as observational constraints
(see Table~\ref{tab:models}).}
\label{h2o-massloss}
\end{figure}

\begin{figure}
\resizebox{\hsize}{!}{\includegraphics{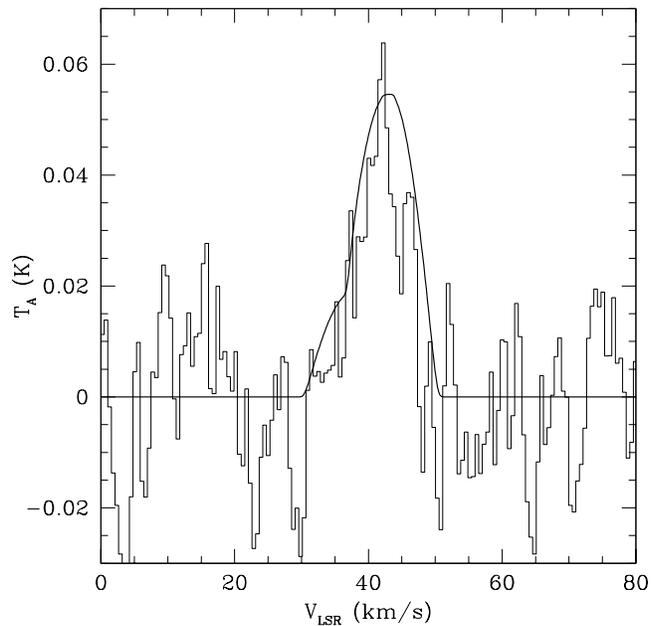}}
\caption{A comparison between the Odin H$_2$O 1$_{10}-1_{01}$ observations
(histogram) and the model profile from the best-fit model. The 
model profile shows clearly the self-absorbed blue wing.
}
\label{modfit}
\end{figure}

\begin{figure*}[t]
\centerline{\includegraphics[angle=-90,width=13.0cm]{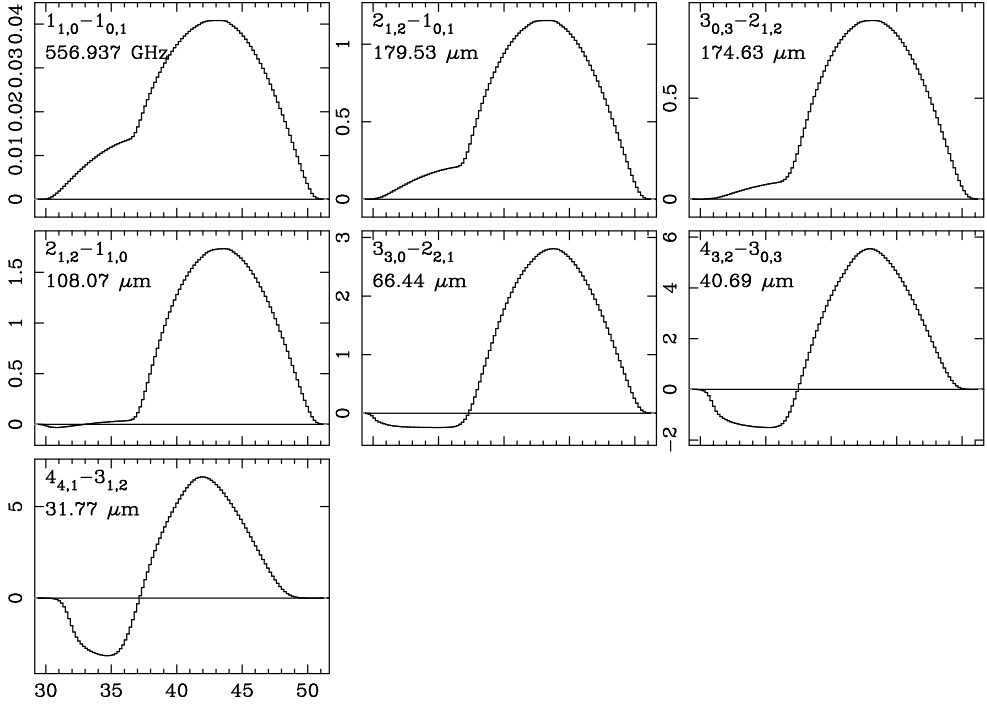}}
\caption{The calculated line profiles for ortho-water from the best-fit model.
The x-axis is velocity in km s$^{-1}$ and the y-axis is flux in the unit
of 10$^{-20}$ W cm$^{-2}$ (km s$^{-1}$)$^{-1}$. The numbers in each
box indicate the transition and the corresponding wavelength.
}
\label{o-h2o}
\end{figure*}

The results of the line emission modelling
are given in Fig.~\ref{h2o-massloss} in terms of the minimum $\chi^2$
for a given combination of ortho-H$_2$O abundance and mass-loss
rate. Clearly, the circumstellar H$_2$O line emission suggests a
relatively low mass-loss rate. Using a 1$\sigma$ uncertainty
we infer a mass-loss rate of (2.5$\pm$0.5)\,10$^{-7}$\,M$_{\odot}$\,yr$^{-1}$,
and an ortho-H$_2$O abundance of (1.0$\pm$0.4)\,10$^{-3}$. The
best-fit model line profile for the Odin 557~GHz line is presented in
Fig.~\ref{modfit}, superimposed on the observed spectrum.
The resulting Odin and ISO lines are shown in Fig.~\ref{o-h2o}
although at a much higher spectral resolution than achieved by ISO.

\begin{table}
\caption{Resulting ortho-H$_2$O line fluxes for the best-fit model
with a stellar mass-loss rate of 2\,10$^{-7}$\,M$_{\odot}$\,yr$^{-1}$
and an ortho-H$_2$O abundance of 1.1\,10$^{-3}$. 
The lines listed are the Odin line (538.29\,$\mu$m) and some ISO lines
previously detected by $a$ Barlow et al. 
(\cite{barlow}) and $b$ Neufeld et al. (\cite{neufeld}). See text for details
on the observed line fluxes and their uncertainties.
}
\begin{tabular}{lccc}
\hline \hline
$\lambda$ ($\mu$m) & Instr. &
\multicolumn{2}{c}{Line flux ($10^{-20}$\,W\,cm$^{-2}$)}\\
\hline
             &      &  Obs & Mod \\
\hline 
538.29       & Odin & 0.36$\pm$0.1  & 0.45 \\
179.53$^{a}$ & LWS  & 10.3$\pm$1.7  & 11.6 \\
174.63$^{a}$ & LWS  & 11.9$\pm$1.2  & 8.33 \\
108.07$^{a}$ & LWS  & 22.8$\pm$4.0  & 15.3 \\
66.44$^{a}$  & LWS  & 32.7$\pm$12.1 & 21.5 \\
40.69$^{b}$  & SWS  & 23.4$\pm$5.3  & 32.6 \\
31.77$^{b}$  & SWS  & 36.3$\pm$3.7  & 31.1\\
\hline
\label{tab:models}
\end{tabular}
\end{table}

We have 
also modelled the para-water lines present in the ISO spectrum.
Since the ortho- and para-water 
molecules do not interact with each other, they can be modelled 
independently. In order to fit these lines for the mass-loss
rate derived above, the required
para-water abundance is 1.1\,10$^{-3}$ 
(with an estimated uncertainty similar to that of the ortho-H$_2$O abundance),
The ortho-to-para ratio is close to unity, in agreement with
that reported by Barlow et al. (\cite{barlow}). This ratio
is expected to be three when considering newly formed H$_{2}$
on the grain surface at high temperature (Tielens \& Allamandola 
\cite{tielens}). The ratio drops to zero for a cold gas.

The total (ortho + para) water abundance according to this analysis,
consistent with the observed intensities, is therefore 
(2.0$\pm$1.0)\,10$^{-3}$.

The marked increase in the $\chi^2$ value for higher mass-loss
rates is primarily an effect of the increasing difficulty to fit
the SWS lines.
The model line fluxes are significantly lower than the observed ones
due to the fact that the blue edge of the profile is in
absorption, cancelling out the emission from the red side of the profile
(see Fig~\ref{o-h2o}). 
The absorption is not so pronounced in the low-mass-loss-rate model,
hence the overall line fluxes agree more with the observations.
It is unfortunate that the resolution of the ISO 
lines is not high enough for the lines to be resolved. 
It should be remarked here
that the SWS lines come from the very inner region of the circumstellar
envelope, and our modelling of these lines is more uncertain.
Fitting only the LWS lines gives a weaker constraint on the combination
of the mass-loss rate and water abundance.

We have also investigated the effect of the assumed 
radius of the water envelope by varying this with respect to
that adopted from the OH maser data.  By
increasing the radius by a factor of three,
the resulting line fluxes changed by less than 50\% for our
best-fit model.
The line fluxes are primarily excitation limited and they
depend only weakly on the adopted H$_{2}$O envelope size. 
Netzer \& Knapp (\cite{netzer}) derived a size of the water envelope
from a theoretical model of H$_2$O photodissociation
\begin{equation}
 R_{\rm out} = 1\,10^{15}\,(\dot{M}/10^{-7})^{0.7} \, {\rm cm,}
\label{e:photrad} 
\end{equation}
where the mass-loss rate is given in M$_{\odot}$ yr$^{-1}$. We
find that the radius obtained from the OH maser data
is fully consistent with this relation for the mass-loss rate
of 2.5\,10$^{-7}$\,M$_{\odot}$\,yr$^{-1}$ obtained from the H$_2$O
line flux fitting.

In our calculation, we did not include the effect of radiative
excitation through the vibrational lines. W Hya has a deep H$_{2}$O
absorption band at 3~$\mu$m, due to the stretching mode, as
seen in the ISO SWS spectrum (Justtanont et al. \cite{kay04}).
This band occurs close to the stellar photosphere
($T_{\rm ex} \sim 10^{3}$ K) and most likely a significant
fraction of the stellar flux at the wavelengths of the
vibrational lines is absorbed in it (Ryde et al. \cite{ryde}).
The detection of H$_2$O maser emission from rotational
lines in the $\nu_2$ vibrational state towards W~Hya shows that radiative
excitation to this state occurs (Menten \& Melnick \cite{menten}).
Deguchi \& Rieu (\cite{deguchi})
found that vibrational radiative excitation plays a
minor role when calculating the rotational line fluxes in the ground state.
On the contrary, Troung-Bach et al. (\cite{tbach}), in their
study of the O-rich Mira variable R~Cas, found that
the ground-state line fluxes decreased by about a factor of two when
radiative excitation through the $\nu_2$ vibrational state was removed.
We believe that the uncertainty in the H$_2$O abundance introduced
by omitting radiative excitation through vibrational states is less
than our estimated uncertainty from the line modelling. 
The effect is likely to lower the abundance
somewhat. We note here that Truong-Bach et al. derive an 
H$_2$O abundance for R~Cas which is remarkably low (in terms of our
modelling results), 1\,10$^{-5}$, considering that R~Cas and W~Hya have 
very similar mass-loss rates, luminosity,
distances, and ISO line fluxes.

\section{The mass-loss rate of W Hya}

There is a large discrepancy when it comes to the determination of the
mass-loss rate of W Hya as a wide range (two orders of magnitudes) of 
the mass-loss rate has been reported over the years.
In this paper we have shown that the circumstellar
CO radio lines suggest a low mass-loss rate, about 
2\,10$^{-7}$\,M$_{\odot}$\,yr$^{-1}$. This result has a rather low 
uncertainty provided that the adopted circumstellar model and CO abundance 
(2\,10$^{-4}$) are correct. Our result is also consistent
with those of Wannier \& Sahai (\cite{wannier}), Young (\cite{young}), and
Olofsson et al. (\cite{hans02}). Once these values are scaled to the same 
distance, velocity, and CO abundance, the
range of the mass-loss rates is (5--20)\,10$^{-8}$\,M$_{\odot}$\,yr$^{-1}$.

Previous reports on the mass-loss rates derived from H$_{2}$O line
emission are based on ISO results. The mass-loss rate estimates range
from as high as 10$^{-5}$ (assumed distance 95\,pc and SWS lines; 
Neufeld et al. \cite{neufeld}) to
6\,10$^{-7}$ M$_{\odot}$ yr$^{-1}$  (assumed distance 130\,pc and 
LWS lines; Barlow et al. \cite{barlow}).
Barlow et al. used a water abundance of 8\,10$^{-4}$ for $r \leq 
6\,10^{14}$\,cm and 3\,10$^{-4}$ beyond this radius.
Neufeld et al. do not give their assumed water abundance, but it must have
been low, according to our results  (see Fig.~\ref{h2o-massloss}). 
We find that the Odin and ISO H$_2$O line intensities suggest a
mass-loss rate of 2.5\,10$^{-7}$\,M$_{\odot}$\,yr$^{-1}$, a value
fully consistent with our CO line modelling result.

The OH maser line data give some hints on the mass-loss rate. Recently,
the 1612 MHz line was detected (Etoka et al. \cite{etoka}), and it is 
substantially weaker than the main lines at 1665 and 1667 MHz. This 
is an indication of a low mass-loss rate (Szymczak \& Engels \cite{szym2}).
Likewise, as pointed out above, the measured size of the OH maser envelope is
consistent with a low mass-loss rate.

Justtanont et al. (\cite{kay04}) derived a low dust mass-loss rate 
by fitting the SED. 
Using their derived dust mass-loss rate combined 
with a dynamical calculation of a dust-driven wind, we infer a total 
mass-loss rate of 1.4\,10$^{-7}$ M$_{\odot}$ yr$^{-1}$. 
This value is only a lower limit because the different dust components 
used in the SED modelling do not account for the 13$\mu$m dust feature. 
Once the carrier is found for this feature and taken into account in the 
model, it is possible that the dust mass loss rate can increase by up to 
a factor of two. The indication from the dynamical calculation is 
W Hya has a relatively low mass-loss rate.

However, Zubko
\& Elitzur (\cite{zubko}) reported a rather high mass-loss rate when
combining SED modelling and a dynamical model for the outflow,
2.3\,10$^{-6}$\,M$_{\odot}$\,yr$^{-1}$ (at a distance of 115\,pc).
The high mass-loss rate is because the authors did not  fit the 
10$\mu$m silicate feature which is the key in deriving the dust mass 
loss rate.

Based on these results we conclude that there is strong evidence that
W Hya has a rather low mass-loss rate, of the order a few 
10$^{-7}$\,M$_{\odot}$\,yr$^{-1}$. 

\section{The circumstellar H$_2$O abundance}

In this work, we derive a consistent mass-loss rate from 
both the CO and H$_{2}$O modelling of (2.5$\pm$0.5)\,10$^{-7}$
M$_{\odot}$\,yr$^{-1}$. The observed line fluxes are reasonably
well produced.
Our derived water abundance for this
mass-loss rate is
exceptionally high, (2.0$\pm$1.0)\,10$^{-3}$. Additional uncertainties
in this estimate are provided by the effects of radiative excitation
through vibrational states, and the uncertain values for the collisional
cross sections.

This derived total water abundance violates the limit set by the cosmic 
abundance of atomic oxygen. In the stellar equilibrium chemistry of an
oxygen-rich environment, all the available carbon should be locked
up  in the form of CO, and the rest of the oxygen goes
into the formation of H$_{2}$O molecules, with a very small fraction
bound into grains and other molecules. The cosmic abundance of
Carbon and Oxygen relative to H is 3.6\,10$^{-4}$ and 8.5\,10$^{-4}$,
respectively (Anders \& Grevesse \cite{anders}). 
Therefore, unless there is a separate influx of water from another
source the water abundance relative to H$_{2}$, 
$f_{\rm H_{2}O}$\,=\,[H$_{2}$O]/[H$_{2}$],
cannot exceed 10$^{-3}$.

The extra water mass required to explain the abundance
needed to model the observed water line emission is
\begin{equation}
M_{\rm H_{2}O} = \dot{M} (f_{\rm H_{2}O} - 10^{-3}) (r_{\rm H_{2}O} / v) 
              (m_{\rm H_{2}O}/m_{\rm H_{2}})
\end{equation}
where $r_{\rm H_{2}O}$ is
the radius of the water shell, and $v$ the gas expansion velocity.
The total extra mass needed is (6.1$\pm$3.0)\,10$^{26}$\,g, i.e., 
(0.1$\pm$0.05)\,M$_{\oplus}$.
This is considerably smaller than the estimated mass of water ice in the 
Solar System, assuming most of the present mass of water is bound in 
comets. The current estimate of the original mass of the Oort
Cloud is highly uncertain and varies between
14--1\,000~M$_{\oplus}$ (Bailey \& Stagg \cite{bailey};
Weissman \cite{weissman}; Marochnik et al. \cite{maroch}). We speculate
that icy bodies can be either part of the disk (and planets) left over after 
the formation of the central star, or a remnant envelope of the original
cloud from which the star formed, reminiscent of the Oort Cloud. 
The disk/envelope remained in stable orbits until the star
ascended the AGB. Because of the much higher luminosity of
the AGB star, compared to that in its 
main sequence phase, outgassing from icy bodies will contribute to
the observed high water abundance. Of course, this requires a steady
supply of new water, since photodissociation effectively destroys
the water molecules on a short time scale. A similar hypothesis has been
put forward to explain the presence of gaseous water around the carbon
star IRC+10216 (Melnick et al. \cite{melnick}). 

\section{Summary}

Observations of the ground-state transition of ortho-water, 1$_{10}-1_{01}$, 
in the circumstellar envelope of W Hya were done using the Odin satellite. 
The line is centered at $V_{\rm LSR}$\,=\,42\,km\,s$^{-1}$ and its width 
corresponds to a gas expansion velocity of about 7\,km\,s$^{-1}$. These 
results compare well with those of previous CO observations (e.g., Kerschbaum 
\& Olofsson \cite{franz}), as well as the new CO data presented in this paper. 
The line flux is (3.6$\pm$1.2)\,10$^{-21}$
W cm$^{-2}$, consistent with the SWAS observation of the same line 
(Harwit \& Bergin \cite{harwit}).

To fit the Odin and ISO water lines, the ortho-water abundance and
mass-loss rate are 
required to be (1.0$\pm$0.4)\,10$^{-3}$ and (2.5$\pm$1.0)\,10$^{-7}$
M$_{\odot}$\,yr$^{-1}$, respectively. Similarly, the para-water abundance 
is estimated to be (1.1$\pm$0.8)\,10$^{-3}$ by fitting ISO lines. This 
mass-loss rate is consistent with that derived from our CO line modelling, 
as well as that obtained
by combining SED modelling and a dynamical model of the outflow.
From our calculation, we show that the line shapes of these water
transitions are different. Unfortunately, from ISO, we have only the
information on the line intensities.
It is noted here that the Heterodyne Instrument for the Far-Infrared
(HIFI) aboard the Herschel Observatory which will be launched around 2008
will have a much higher resolution which can resolve some of these ISO lines,
thereby giving another constraint on the model parameters.

If the mass-loss rate for W Hya is as low as 
(2.5$\pm$0.5)\,10$^{-7}$ M$_{\odot}$ yr$^{-1}$, the total 
water abundance needed to explain 
the observed line fluxes is (2.0$\pm$1.0)\,10$^{-3}$.
However, if we assume the cosmic abundance for carbon and oxygen, the 
maximum total water abundance, relative to H$_{2}$,
cannot be higher than 10$^{-3}$. A way to reconcile the high
total water abundance needed to model the H$_{2}$O lines of W Hya
is an extra injection of water in the system, possibly from 
evaporation of icy bodies, such as planets or comets. This is the 
suggested explanation for the detection of water in IRC+10216, the IR-brightest 
carbon-rich AGB star (Melnick et al. \cite{melnick}). The excess 
H$_{2}$O mass needed to explain the Odin and ISO observations in 
W Hya is small, of the order of (0.1$\pm$0.05)\,M$_{\oplus}$,
compared to the estimate of the original mass of the Oort Cloud.

\begin{acknowledgement}

This work is partly funded by the Swedish National Space Board.
We wish to thank Ren\'{e} Liseau and G\"{o}ran Olofsson for
fruitful discussions. Our research made use of the database SIMBAD.

\end{acknowledgement}


\begin{thebibliography}{}

\bibitem [1989]{anders}  Anders E., \& Grevesse N., 1989, Geochemica et
                         Cosmochimica Acta 53, 197
\bibitem [1988]{bailey}  Bailey M.E., \& Stagg C.R., 1988 MNRAS 235, 1
\bibitem [1996]{barlow}  Barlow M.J., Nguyen-Q-Rieu, Truong-Bach et al.,
                         1996, A\&A 315, L241
\bibitem [2002]{cami}    Cami J., 2002, PhD thesis, University of Leiden
\bibitem [1997]{cerni}   Cernicharo J., Alcolea J., Baudry A., Gonz{\'a}lez-Alfonso
                         E., 1997, A\&A 319, 607
\bibitem [1994]{chapman} Chapman J.M., Sivagnanam P., Cohen R.J., Le Squeren
                         A.M., 1994, MNRAS 268, 475
\bibitem [1990]{deguchi} Deguchi T., \& Rieu N.-Q., 1990, ApJ 360, L27
\bibitem [1997]{esa}     ESA, 1997, the Hipparcos and Tycho Catalogues, ESA
                         SP-1200
\bibitem[2003]{etoka}    Etoka S., Le Squeren A.M., Gerard E. 2003, A\&A 403, L51
\bibitem [2001]{flower}  Flower D.R., 2001, J. Phys. B.: At. Mol. Opt. Phys. 34,
                         2731
\bibitem [2003]{frisk}   Frisk U., Hagstr\"{o}m M., Ala-Laurinaho J., et al., 
                         2003, A\&A 402, L27
\bibitem [1969]{gilman}  Gilman R.C., 1969, ApJ 155, L185
\bibitem [1993]{green}   Green S., Maluendes S., McLean A.D., 1993, ApJS 85, 181
\bibitem [1996]{habing}  Habing H.J., 1996, A\&ARv 7, 97
\bibitem [1995]{haniff}  Haniff C.A., Scholz M., Tuthill P.G., 1995, MNRAS
                         276, 640
\bibitem [2002]{harwit}  Harwit M., \& Bergin E.A., 2002, ApJ 565, L105
\bibitem [1990]{hawkins} Hawkins G.W., 2000, A\&A 229, L5
\bibitem [2003]{ake}     Hjalmarson \AA, Frisk U., Olberg M., et al., 2003, 
                         A\&A 402, L39
\bibitem [1993]{iben}    Iben Jr., I., \&, Renzini A., 1983 Ann Rev A\&A 21, 271
\bibitem [2004]{kay04}   Justtanont K., de Jong T., Tielens A.G.G.M., 
                         Feuchtgruber H., Waters L.B.F.M., 2004, A\&A 417, 625
\bibitem [1999]{franz}   Kerschbaum F., Olofsson H., 1999, A\&AS 138, 299
\bibitem [2003]{knapp}   Knapp G.R., Pourbaix D., Platais I., Jorissen A., 2003,
                         A\&A 403, 993
\bibitem [2000]{leb}     Lebzelter T., Kiss L.L., Hinkle K.H., 2000 A\&A 361,
                         167
\bibitem [1993]{loup}    Loup C., Forveille T., Omont A., Paul J.F., 1993, 
                         A\&AS 99, 291
\bibitem [1988]{mamon}   Mamon G.A., Glassgold A.E., Huggins P.J., 1988 ApJ 328, 797
\bibitem [1988]{maroch}  Marochnik L.S., Mukhim L.M., Sagdeev R.Z., 1988,
                         Science 242, 547
\bibitem [2001]{melnick} Melnick G.J., Neufeld D.A., Ford K.E.S., Hollenbach 
                         D.J., Ashby M.L.N., 2001, Nature 412, 160
\bibitem [1989]{menten}  Menten K.M., Melnick G.J., 1989, ApJ 341, L91
\bibitem[2003]{millar}   Millar T.J.. 2003, in Asymptotic giant branch stars, 
                         Astronomy and astrophysics library, New York, Berlin: 
                         Springer, ed. H.~J. {Habing} \& H.~{Olofsson}, p.248  
\bibitem [1987]{netzer}  Netzer N., \& Knapp G.R., 1987, ApJ 323, 734
\bibitem [1996]{neufeld} Neufeld D.A.,  Chen W., Melnick G.J. et al., 1996,
                         A\&A 315, L237
\bibitem [2003]{nordh}   Nordh H.L., von Sch\'{e}ele F., Frisk U., et al.
                         2003, A\&A 402, L21
\bibitem [2004]{ohnaka}  Ohnaka K., 2004, A\&A 424, 1011
\bibitem [2003]{olof03}  Olofsson H. 2003, in Asymptotic giant branch stars, 
                         Astronomy and astrophysics library, New York, Berlin: 
                         Springer, ed. H.~J. {Habing} \& H.~{Olofsson}, p.325
\bibitem [2002]{hans02}  Olofsson H., Gonz\'{a}lez-Delgado D., Kerschbaum F., 
                         Sch\"{o}ier F. L., 2002, A\&A 391, 1053
\bibitem [2004]{pardo}   Pardo J.R., Alcolea J., Bujarrabal V., et al.,
                         2004, A\&A 424, 145
\bibitem [1996]{phillips}Phillips T.R., Maluendes S., Green S., 1996, ApJS 107, 467
\bibitem [1991]{rybicki} Rybicki G.B., \& Hummer D.G., 1991, A\&A 245, 171
\bibitem [2002]{ryde}    Ryde N., Lambert D.~L., Richter M.~J., Lacy J.~H. 2002,
                         ApJ 580, 447
\bibitem [2001]{fredrik} Sch\"{o}ier F.L., \& Olofsson H., 2001, A\&A 368, 969
\bibitem [2005]{schoeier05} Sch\"{o}ier F.L., van der Tak F.F.S, van Dishoeck, 
                         E.F., Black J.H., 2005, A\&A 432, 369
\bibitem [1998]{szym}    Szymczak M., Cohen R.J., Richards A.M.S., 1998,
                         MNRAS 297, 1151
\bibitem [1998]{szym2}   Szymczak M. \& Engels D., 1995, A\&A 296, 727
\bibitem [1987]{tielens} Tielens A.G.G.M., \& Allamandola L.J, 1987,
                         In Interstellar Processes, eds. D.J. Hollenbach,
                         H.A. Thronson, Reidel, Dordrecht, p. 397
\bibitem [1974]{treffers}Treffers R, \& Cohen M., 1974, ApJ 188, 545
\bibitem [1999]{tbach}   Truong-Bach, Sylvester R.J., Barlow M.J., et al., 1999,
                         A\&A 345, 925
\bibitem [1986]{wannier} Wannier P.G., \& Sahai R., 1986, ApJ 311, 335
\bibitem [1991]{weissman}Weissman P.R., 1991, Icarus 89, 190
\bibitem [1995]{young}   Young K., 1995, ApJ 445, 872
\bibitem [2000]{zubko}   Zubko V., \& Elitzur M., 2000, ApJ 554, L137
\end{thebibliography}
\end{document}